\documentclass[prb,aps,twocolumn]{revtex4-1}

\usepackage{graphicx}

\begin{document}

\title{Universal electric-field-driven resistive transition in narrow-gap Mott insulators}
\author{P. Stoliar,$^{1,2}$ L. Cario,$^{3}$ E. Janod,$^{3}$ B. Corraze,$^{3}$C. Guillot-Deudon,$^{3}$ S. Salmon-Bourmand,$^{3}$ V. Guiot,$^{3}$ J. Tranchant,$^{3}$ M. Rozenberg$^{1,\ast}$}
\affiliation{$^{1}$Laboratoire de Physique des Solides, CNRS UMR 8502, Universit\'{e} Paris Sud, B\^{a}t 510, 91405 Orsay, France}
\affiliation{$^{2}$ECyT, Universidad Nacional de San Mart\'{i}n, Campus Miguelete, 1650 San Mart\'{i}n, Argentina}
\affiliation{$^{3}$Institut des Mat\'{e}riaux Jean Rouxel (IMN), Universit\'{e} de Nantes, CNRS, 2 rue de la Houssini\`{e}re, BP32229, 44322 Nantes, France}

\maketitle

\global\long\def\si{Supporting information}

\global\long\def\PCMMI{P_{{\rm CM\to MI}}}
\global\long\def\PMICM{P_{{\rm MI\to CM}}}
\global\long\def\RCM{R_{{\rm CM}}}
\global\long\def\RMI{R_{{\rm MI}}}
\global\long\def\NCM{N_{{\rm CM}}}
\global\long\def\NMI{N_{{\rm MI}}}

\global\long\def\EB{E_{B}}
\global\long\def\ECM{E_{{\rm CM}}}
\global\long\def\Eth{E_{th}}
\global\long\def\Vth{V_{th}}
\global\long\def\td{t_{d}}
\global\long\def\RL{R_{L}}
\global\long\def\trelax{t_{relax}}

\global\long\def\GaTaSe{GaTa$_{4}$Se$_{8}$}
\global\long\def\NiS{NiS$_{2-x}$Se$_{x}$}
\global\long\def\VO{V$_{2-x}$Cr$_{x}$O$_{3}$}
\global\long\def\AMX{AM$_{4}$Q$_{8}$}

One of today's most exciting research frontier and challenge
in condensed matter physics is known as Mottronics, whose
goal is to incorporate strong correlation effects into the realm
of electronics. \cite{Inoue2008,Hwang2012} In fact, taming the Mott insulator-to-metal
transition (IMT), which is driven by strong electronic correlation
effects, holds the promise of a commutation speed set by
a quantum transition, and with negligible power dissipation.\cite{itrs2011}
In this context, one possible route to control the Mott transition
is to electrostatically dope the systems using strong dielectrics,
in FET-like devices.\cite{Takagi2010,Nakano2012,Asanuma2010} Another possibility is through resistive
switching,\cite{Rozenberg2011} that is, to induce the insulator-to-metal transition
by strong electric pulsing.\cite{itrs2011,Cario2010,Souchier2011PSSR,Tranchant2013,Vaju2008MEE} This action brings the correlated
system far from equilibrium, rendering the exact treatment
of the problem a difficult challenge. Here, we show that
existing theoretical predictions of the off-equilibrium manybody
problem \cite{Oka2010,Heidrich-Meisner2010} err by orders of magnitudes, when compared
to experiments that we performed on three prototypical
narrow gap Mott systems \VO, \NiS\  and  \GaTaSe,
and which also demonstrate a striking universality of this Mott
resistive transition (MRT). We then introduce and numerically
study a model based on key theoretically known physical features
of the Mott phenomenon in the Hubbard model.\cite{Georges1996} We
find that our model predictions are in very good agreement
with the observed universal MRT and with a non-trivial timedelay
electric pulsing experiment, which we also report. Our
study demonstrates that the MRT can be associated to a dynamically
directed avalanche.

\begin{figure*}
\includegraphics[width=17cm]{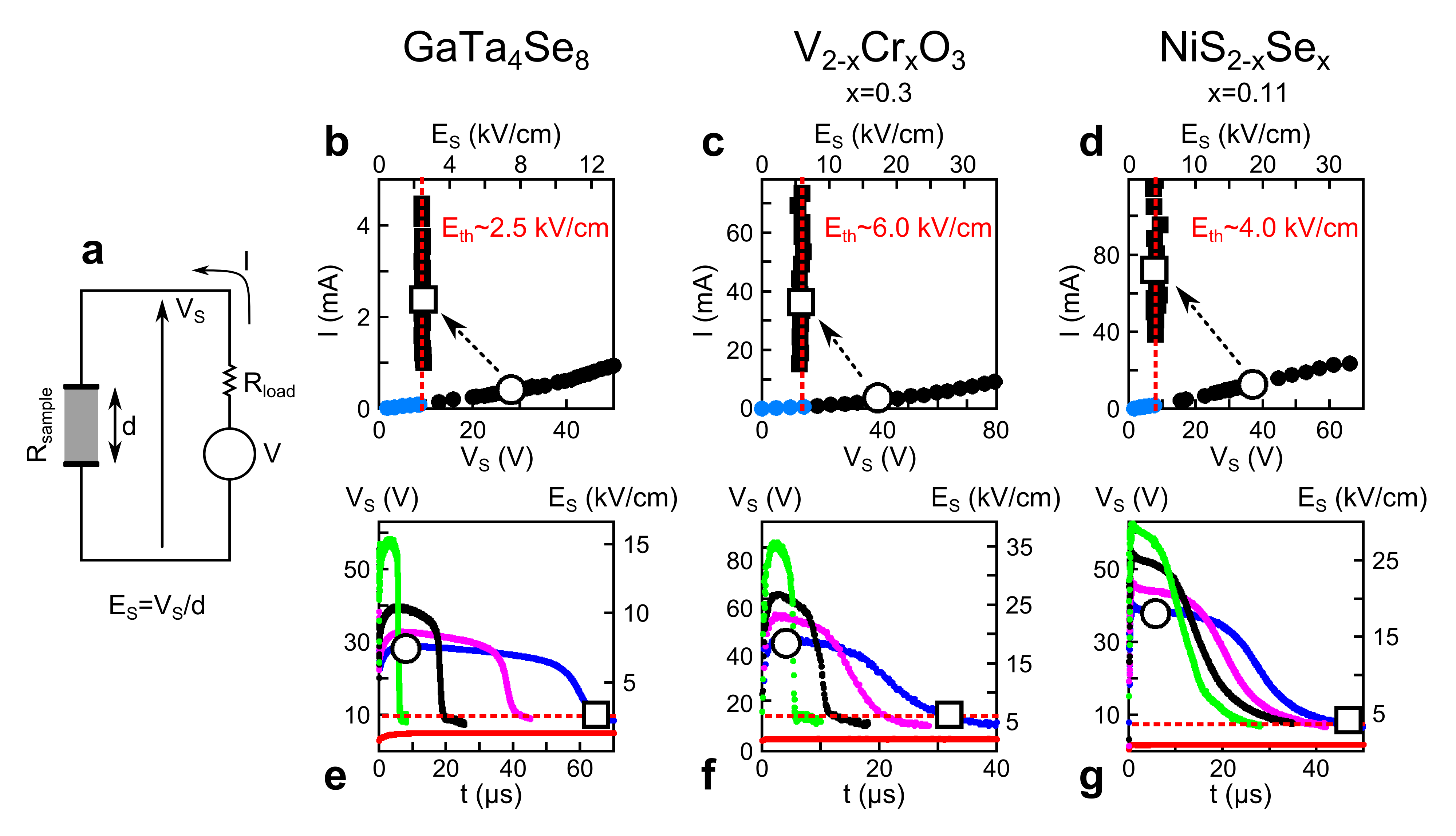}
\caption{a) The schematics of the
experimental setup. b-g) Universal $I-V$ characteristics (panels b, d, f) and time dependence of the
sample voltage $V_S(t)$ (panels c, e, g) for three different types of narrow gap Mott insulators.
Blue dots correspond to the region below $E_{th}$, where no breakdown is observed.
Black symbols correspond to the $I-V$ characteristic in the resistive switching region, above $E_{th}$. 
The black dots show the initial $I-V$, before the breakdown, and the black squares indicate the final state.
The open symbols highlight a particular breakdown transition for easier visualization. 
Measurements on \GaTaSe were performed at 77 K, on V$_{2-x}$Cr$_x$O$_3$ (x=0.3) at 164 K and on NiSe$_{2-x}$S$_x$ (x=0.11) at 4 K.
For further experimental details see the Supporting Information.}
\end{figure*}

The resistive transition in Mott insulators is a topic of great
current interest in both basic and applied condensed matter
physics. On one hand, the understanding of the behavior of
a strongly correlated system far from equilibrium remains
a formidable many-body physics problem.\cite{Oka2010,Heidrich-Meisner2010,Oka2003,Oka2005,Taguchi2000,Dutta2007,Eckstein2010} While on the
other hand, it is also a key issue for Mottronics,\cite{Inoue2008} (i.e., the current
challenge of bringing strong correlation effects into the
realm of electronics).\cite{Hwang2012}

The theoretical work on the electric-field-driven breakdown
of a Mott state in the Hubbard model is consistent with an intuitive
scenario for the onset of the transition.\cite{Oka2010,Heidrich-Meisner2010,Oka2003,Oka2005,Ahn2006} In fact, the
onset of electric conduction is predicted to occur for a strength
of the electric field $E_{th}$ of the order of the Mott-Hubbard gap $\Delta$ divided by a length $\xi$ of the order of the unit cell. In a Mott
insulator the doublon state, which is the virtual double occupation
of a site, is associated to the superexchange interaction,
thus extends within distances $ \sim \xi$. Then, the electric breakdown
due to the deconfinement of the doublon will occur when the
field is such that it bends the Hubbard bands by the gap energy
$\Delta$ within the length $\xi$. Hence, $E_{th} \sim \Delta / \xi$, so the predicted order
of magnitude of the critical field is the order of few V/nm. This
is off by several orders of magnitude from the experimental
observations, which is typically in the kV/cm (i.e., 10$^{-4}$ V/nm)
range.\cite{Taguchi2000,Vaju2008ADMA,Sabeth2012} In addition, and consistent with that observation,
the predicted delay time for the transition, that is the time
elapsed from the application of the external field till the onset
of the conduction jump, is also off with respect to the experimental
findings by about three orders of magnitude. \cite{Cario2010,Souchier2011PSSR,Tranchant2013,Vaju2008MEE,Taguchi2000,Vaju2008ADMA}

These discrepancies signal a shortcoming of current theoretical
approaches to describe the breakdown of Mott insulators
by a simple Hubbard model. It is likely, that additional degrees
of freedom, such as charge or lattice, may also play a significant
role.\cite{Taguchi2000} Evidently, that would add another layer of complexity
to the already difficult out-of-equilibrium treatment of the
problem. Hence, here we take another approach and shall first
show that three different prototypical narrow gap Mott insulators,
which all have a pressure driven metal-insulator transition,
undergo a universal Mott resistive transition phenomenon.
Specifically, they display threshold fields and time delays of
similar magnitudes, which are significantly off from theoretical
estimates. This surprising finding motivates us to introduce a
resistor network model that incorporates basic features known
from equilibrium many-body studies of the Mott transition.\cite{Georges1996}
A similar method was recently used to study current oscillations
in VO$_2$.\cite{Driscoll2012} We study the model by numerical simulations and
show that it captures the qualitative behavior observed in experiments.
In particular, we find: i) that the MRT may be viewed as
a dynamically directed avalanche, and ii) that a new time scale
emerges from the local recovery of inhomogeneously transited
regions. These observations are further validated by comparison
of non-trivial theoretical predictions with experiments of
multiple electric pulsing, which we also verify and report.

We study the electric-field-driven MRT in the following
selected prototypical three dimensional Mott insulators,\cite{Imada1998}, \GaTaSe, \NiS   and \VO, a ternary and a binary
chalcogenide, and an oxide. These compounds are archetypal,
as they all undergo an insulator to metal transition driven by
pressure,\cite{Imada1998,Pocha2005,Dorolti2010,guiot2011,Miyasaka,TaPhuoc2013} which is in good qualitative agreement with the
theoretical predictions of Hubbard model studies in dynamical
mean field theory.\cite{Georges1996,Kotliar2004} The first compound is actually a representative
of a whole family of Mott systems, the \AMX, with A=Ge, Ga, M=Ta, Nb, V, Mo and Q=Se, S.\cite{Pocha2005} Notice that
an electric-field driven transition has also been reported in
VO$_2$. However, unlike the Mott insulators in the present study
does not have a pressure induced transition, but a first order
insulator to metal transition driven by increasing temperature.
\cite{Driscoll2012,Kim2010,Kim2004,Guiot2013,Zimmers2013} This is qualitatively different to the present case,
since the resistivity of our systems merely shows thermal activated
behavior with small gaps of the order of tenths of eV.
In Figure 1 we present the $I - V$ curves of the measured resistive
transitions of the three Mott systems. They all show similar
behavior with a sharp transition onset at an electric field
threshold $\Eth$. The data also show the delay times for the transition
$\td$. As mentioned before, the magnitudes of these quantities
are off by at least three orders of magnitude from estimates
obtained from solutions of Hubbard models.\cite{Oka2010,Heidrich-Meisner2010,Oka2003,Oka2005} We should
also emphasize that the sharp transition is indicative of its
electronic origin. In fact, a reduction of resistance due to Joule
self-heating would produce and S-shaped $I - V$ characteristic
with a gradual evolution of the resistance change.\cite{Ridley1963}

\begin{figure}
\includegraphics[width=8.5cm]{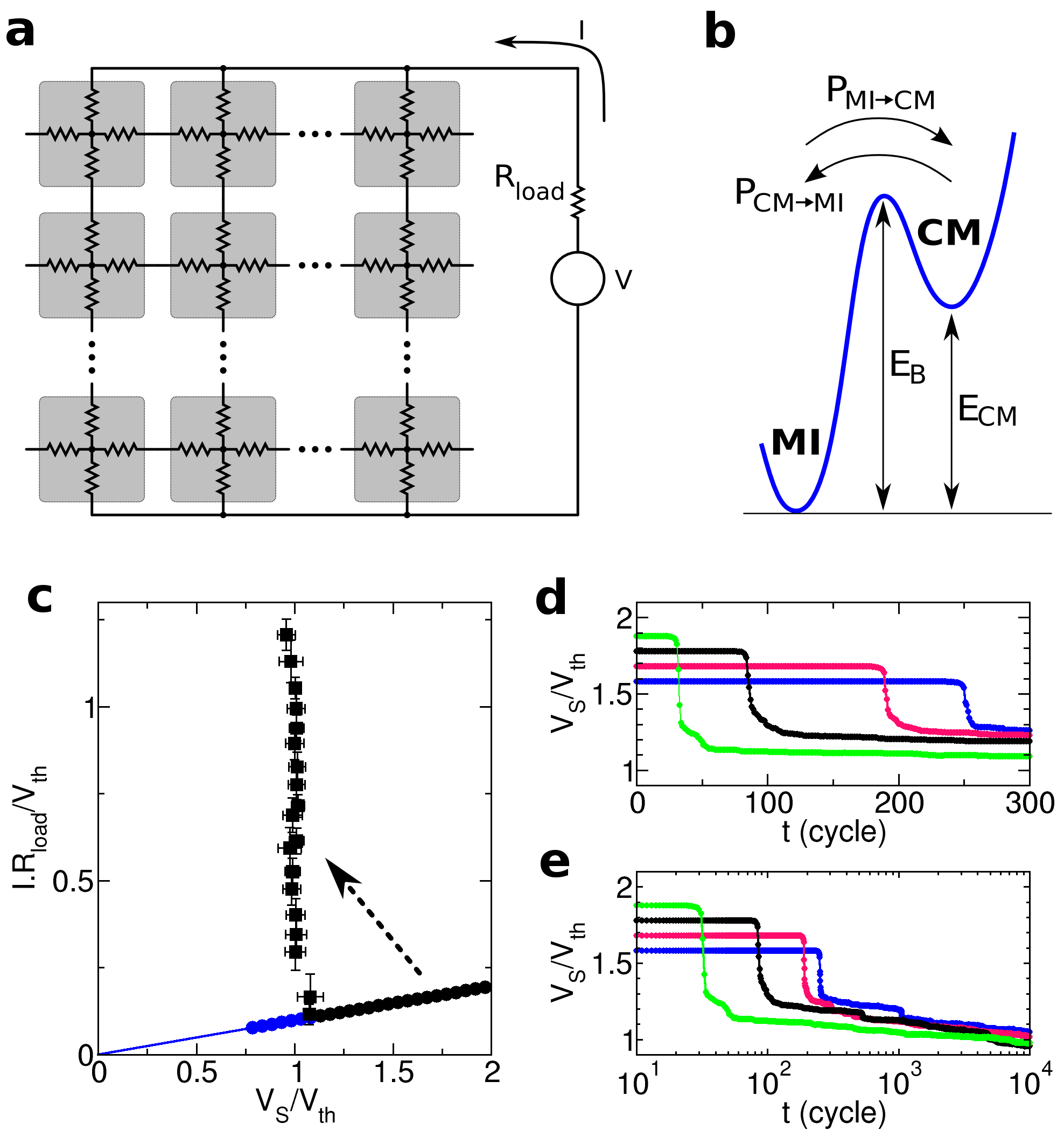}
\caption{a) Mott resistor network model and simulation circuit, in analogy with the experimental setup. 
b) Energy diagram of the CM and MI states and the energy barrier. 
c) Simulation data of the $I$-$V$ characteristics, following the same convention as
in Fig. 1. 
d) Time dependence of the sample voltage as a function of simulation time.
The data correspond to $V_S$/$V_{th}$ = 1.74, 1.84, 1.95 and 2.06, from right to left.
e) Same as panel (d) on semi-log scale to observe the long time behavior.}
\end{figure}

\begin{figure*}
\includegraphics[width=16cm]{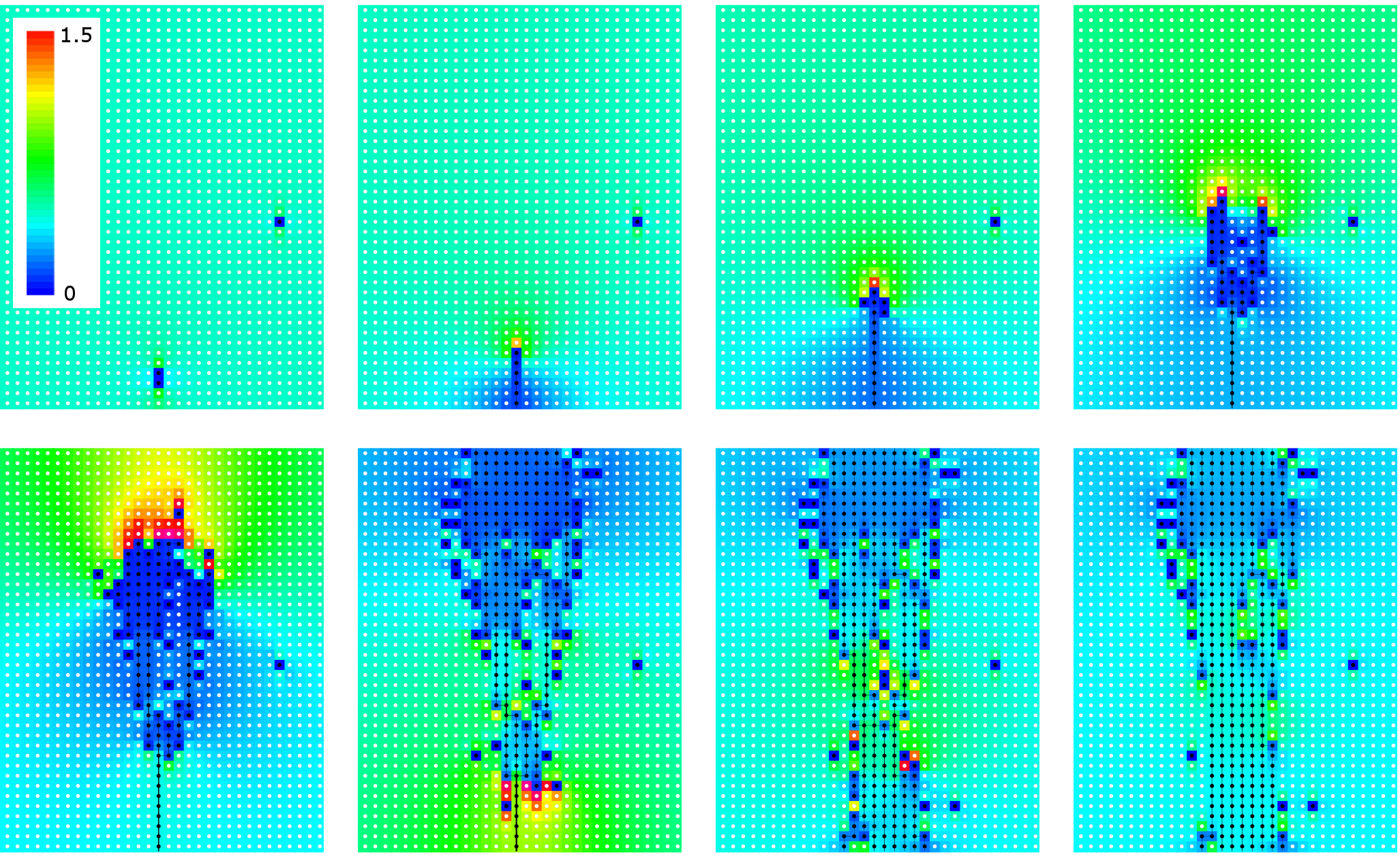}
\caption{Intensity color plot of local voltage drops on the resistor network $\Delta V / E_B$.
Only the region where the filament forms and the breakdown occurs is shown.
The electrodes are at the top and bottom of the panels.
The applied voltage is $V_S$/$V_{th}$ = 1.74, the panes correspond to simulation time $t$ =
60, 70, 75, 80, 83, 85, 87 and 100 (See Supporting
Information for the animated data).
The state of each cell is represented by a white or a black dot, for MI and CM, respectively.}
\end{figure*}

The resistor network model is schematically shown in
Figure 2, where each one of the cells represents a small region
of the physical system, which is in either of two well defined
electronic states, Mott insulator (MI) or correlated metal (CM).
Since the compounds are normally insulators, we assume that
the MI-state is lower in energy, which we define as the reference
$E_{\rm MI} \equiv 0$. The CM-state is assumed to be a metastable state,
with a higher energy $\ECM$, and separated from the MI by an
energy barrier $\EB$ (see Figure 2). These features are based on
previous studies of the Mott-Hubbard transition,\cite{Georges1996} that have
successfully captured many (equilibrium) properties of this
transition.\cite{Imada1998,Limelette2003} Indeed, the dynamical mean field theory studies
on the Hubbard model predict the existence of two competing
states, very close in energy, which have a region of coexistence
close to the first order Mott metal-insulator transition.\cite{Georges1996}
In our model, the MI and CM states are associated with corresponding
high and low resistance values, $\RMI$ and $\RCM$. The
cells are assumed to represent patches of at least a few nanometers
in size, such that an electronic state may be well defined.
Initially all cells are assumed to be in the MI-state. The time is
discretized and the external voltage is applied to the MRN at
each timestep. The local voltage drops $\Delta V({\bf r})$ are then computed
by standard methods.\cite{Kirkpatrick1973} We assume that the MI $\to$ CM transition
rate of each cell is given by $\PMICM=\nu\, e^{-(\EB-q\Delta V)/kT}$,
where the constant $\nu$ is an attempt rate, which we set to unity
and similarly the charge $q$. The precise origin of this transition
is not yet fully established. It may be likely driven by
injected carrier doping, due to impact ionisation, which could
induce the local metalization of a small region.\cite{Kim2004,Guiot2013} However,
other mechanisms including electromechanical coupling
cannot be ruled out,\cite{Dubost2009} as Mott insulators may have significant compressibility anomalies near the metal-insulator transition.
After a MI $\to$ CM transition, the metastable CM-state
may relax back to the MI by overcoming the energy barrier
$\EB$, with a transition rate given by $\PCMMI=\nu\, e^{-(\EB-\ECM)/kT}$.
Notice that, since $\RMI\gg \RCM$, the voltage drop may be
neglected in the CM cells. This relaxation process defines an
important time scale in the problem, as we shall later see.
Under these assumptions, the model can be solved by numerical
simulations in discretized time (see Supporting Information
for details). In analogy to the experimental setup (see
Figure 1), we apply a voltage to the network $V_S=[R_S/(R_S+R_L)]V$, where $V$ is the external voltage, $R_S$ is the equivalent resistance
of the network and $\RL$ is a limiting load resistance. Initially, all
the cells are in the MI state, and $R_S = R_{\rm MI}^0 = g~ \RMI$, where $g$ is a
geometrical factor of the order of unity. Then,
during the simulation, the sites may switch
between the MI and CM states, and thus the
value of $R_S$ is recomputed at every timestep.
The specific values of the adopted parameters
are detailed in the Supporting information.

\begin{figure}
\centering
\includegraphics[width=8.0cm]{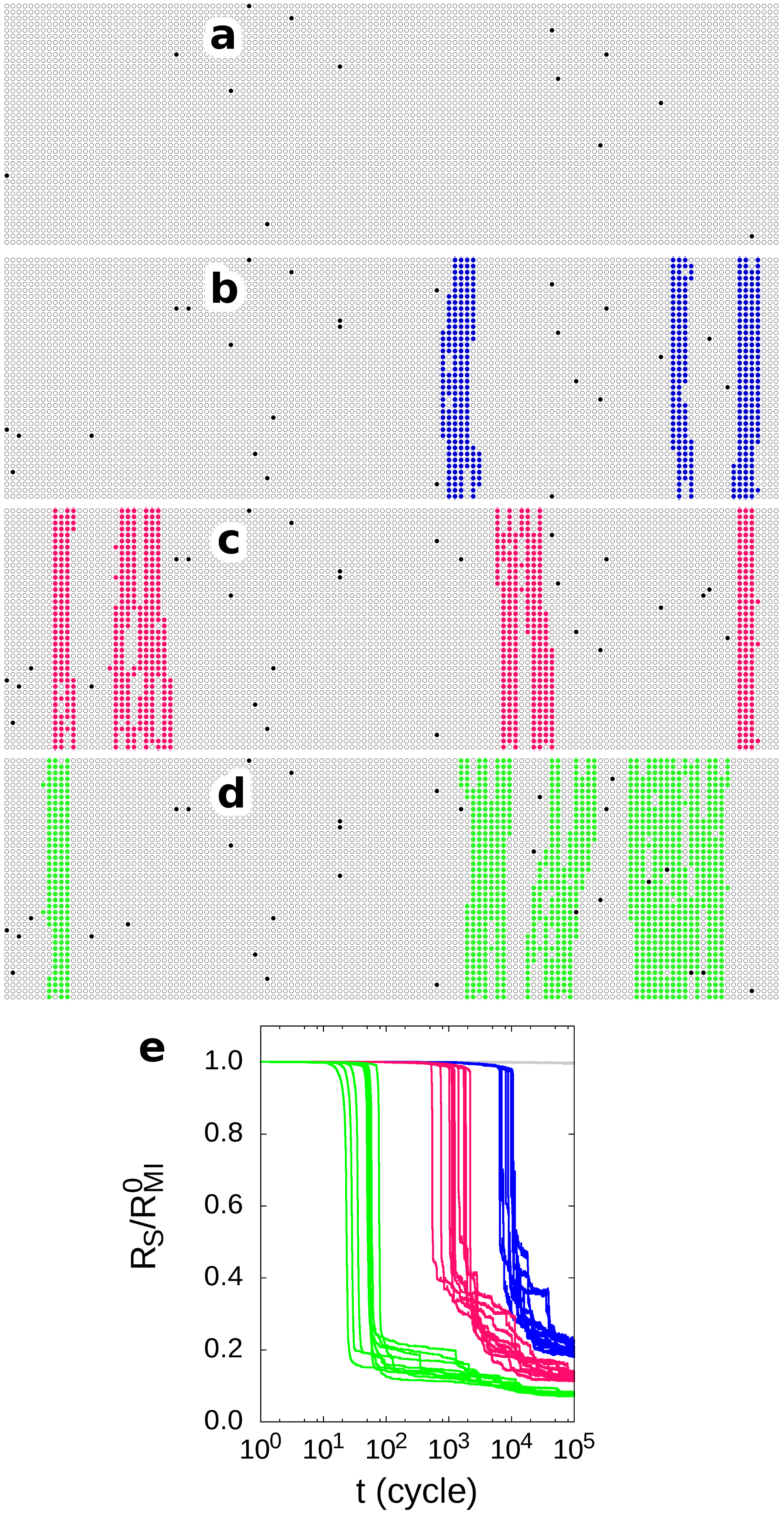}
\caption{Local resistance snapshot of the network after a long simulation time of 10$^{5}$ cycles.
The black and color dots indicate the sites that are in the CM state, while the grey dots correspond to the
sites in the MI state. The dots forming filaments are coloured for easier visualization.
The initially applied sample voltage were $V_S$/V$_{th}$ = 0.7, 1.1, 1.4 and 1.8 (panels (a) to (d)).
The right panel (e) shows the dependence of the normalized network resistance ($R_S / R_{\rm MI}^0$),
as a function of simulation time. The color notation is as in the previous panels.
The dispersion of the data corresponds to 10 runs with different seeds, which illustrate the 
effect of the finite size of the network.}
\end{figure}

As shown in Figure 2, the simulations predict
a sharp threshold behavior for the MRT
phenomenon and $I - V$ characteristics that are
in excellent agreement with the experimental
data of Figure 1. Moreover, the simulations
also capture the qualitative behavior of the
time delay $\td$, which increases dramatically
as the voltage approaches the threshold from
above, and eventually diverges as $V_s \to \Vth$.

While one may expect that the sharp
threshold stems from a simple percolation process,
in fact the simulations unveil a qualitatively
different scenario. At low applied $V$, after
an initial transient, the fraction of sites in the
CM to those in the MI state stabilizes and is
simply given by the ratio between the transition
rates $\PMICM/\PCMMI$, which is small since
$(\EB-\ECM)<(\EB-\Delta V)$. Beneath the threshold,
$\Delta V\approx V/M$, where $M$ is the number of cell
layers between the electrodes (see Figure 2).
This reflects the equilibrium of concentration
condition $\NMI\PMICM = \NCM\PCMMI$.

In a simple percolation picture, the 2D
(site) percolation fraction $\sim$0.59 would be
reached when the applied voltage is such
that $\Delta V\approx \ECM$.\cite{Stauffer1992} However, in our model,
already for voltages well beneath that
value the MRT takes place in a dynamically
directed avalanche phenomenon. The
reason may be simply understood. At low
$\PMICM$ rates (i.e., low applied $V$), the CM
sites are diluted and randomly distributed.
However, as $V$ is increased the production rate of CM sites
grows, and eventually there is a significant chance of having a
few consecutive CM sites longitudinally aligned, in a direction
perpendicular to the electrodes. The voltage drop in that low-$\RCM$ metallic segment decreases, and induces a compensating
increase of the voltage drop at the high-$\RMI$ sites that are close
to its extremes. Thus, the transition probability MI $\to$ CM
for those sites increase as well, which quickly get added to
the metallic segment and further increase the voltage drop
at the extremes, eventually leading to a runaway process, or
avalanche. This is the dynamically generated directed percolation
leading to the MRT. This is visualized in detail in
Figure 3, which shows the formation of a conductive path
and the enhanced transition probability that develops along
the longitudinal direction. Interestingly, this picture is key to
understanding many experimental features, and leads to further
predictions that we shall test later. In Figure 4, we show
the snapshots of the state of the system after a long period of a
continuously applied $V$. Beneath the $\Vth$ the production of CM
sites remains very diluted and far from percolation, with no
precursors. From the observation of the panels corresponding
to $V_S$ above and below $\Vth$, it is clear that the transition occurs
for a fraction of CM sites, which is much smaller than the
2D percolation value mentioned before. Above the threshold,
quite surprisingly, we observe that the number of conductive
paths (or the fraction of the system that is in transited state)
increases with the voltage difference to the threshold, $V_S - \Vth$.
We trace this effect to a rapid succession of multiple avalanche
events. In fact, after the initial path is formed, so long as the
voltage on the sample remains above $\Vth$, the avalanche events
will continue until the applied voltage on the sample decays
down to the value $\Vth$ (see Figure 2e).

Another interesting prediction of the model is that the low
resistance state is not permanent. Since the CM states are
metastable, after the applied voltage is switched off, all those
sites begin to relax back to the MI state. Thus, Mott resistive
transition is, unlike, the permanent dielectric breakdown
phenomenon in semiconductors, a volatile effect (i.e., the
resistance recovers its original value after the applied voltage
is switched off) with a characteristic relaxation time $t_{relax}$
Indeed, experiments reported in the \GaTaSe\ compound
show this sudden recovery effect.\cite{Cario2010,Souchier2011PSSR,Tranchant2013,Vaju2008MEE} Therefore, the Mott resistive
transition phenomenon is a volatile and fully reversible
resistive switching, which is qualitatively different to the nonvolatile
resistive switching effect in transition metal oxide
systems.\cite{Rozenberg2011,Sawa2008,Waser2009,Pershin2011}

\begin{figure}
\includegraphics[width=8.5cm]{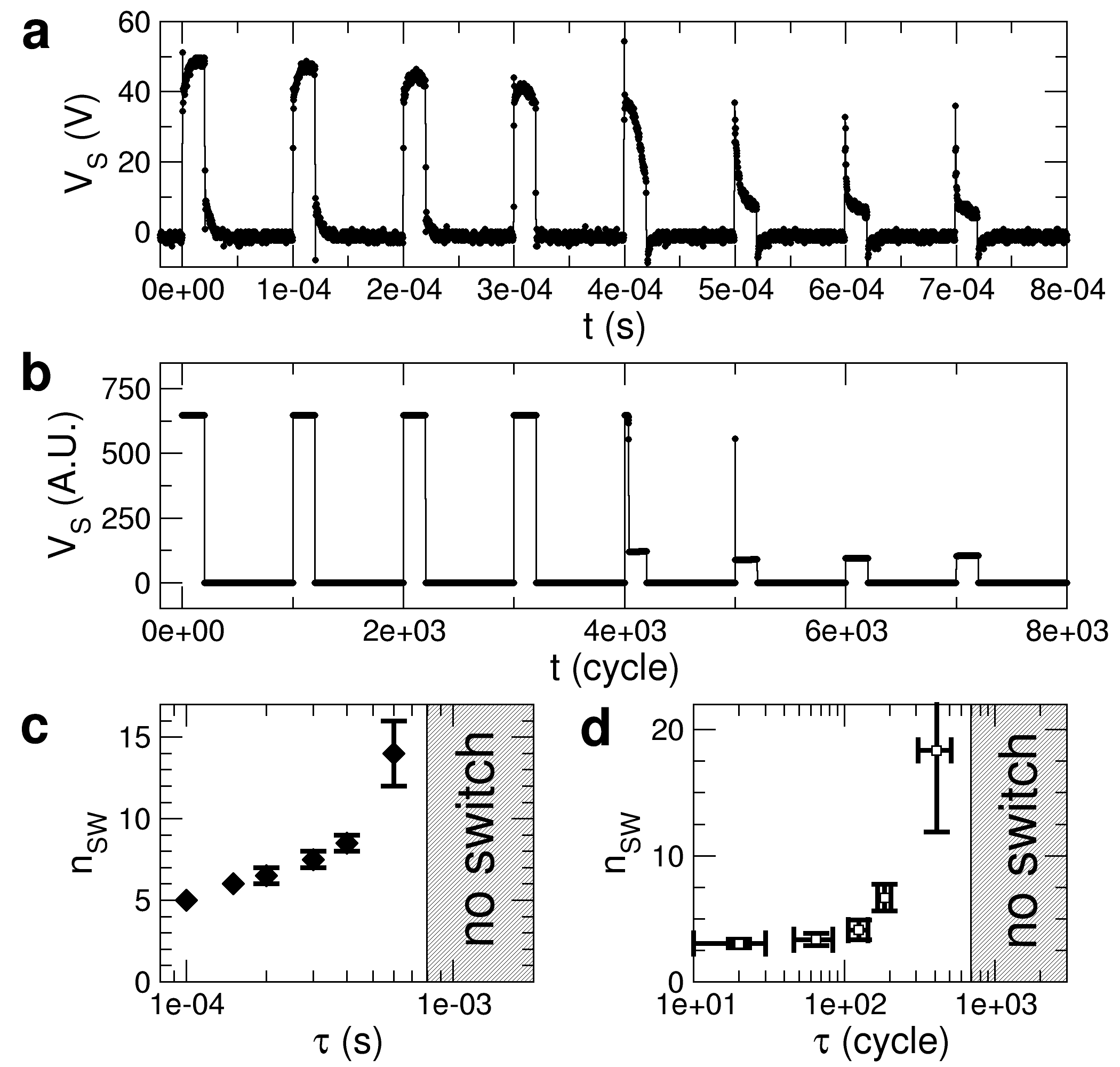}
\caption{Electric pulsing with varying pulse-interval time.
a) Sample voltage $V_S$ as a function of time of a \GaTaSe-based device 
upon the application of a train of voltage pulses of 60 V and duration of 
20 $\mu$s with a period $\tau$=100 $\mu$s. 
Notice that for a continuously applied voltage of 60 V, the device switches after 
a delay time of $\td=70\mu$s (see Supporting Information). 
b) Simulation data for a train pulse showing the same qualitative behavior.  
c) Experimental number of pulses required for switching as a function of the pulse period. The
pulse duration is as in (a). The grey area corresponds to a pulse-interval where switching
no longer occurs. It indicates that the $\trelax\sim800\mu s$.
d) Simulation results for varying the pulse-interval.
Error bars are due to finite size effects.}
\end{figure}

We finally consider a non-trivial prediction, which follows
from the assumptions of our model. Since, one of its key features
is relaxation of the CM cells back to the MI state, we may
expect a qualitative different behavior if we apply short voltage
pulses (i.e., of a duration smaller that the respective $t_d$), which
are repeated at a time interval that may be either shorter or
longer than $\trelax$. In fact, since a single pulse is too short to produce
the transition, in the case of long time intervals between
pulses the effect of each one vanishes due the relaxation before
the next pulse arrives. In contrast, if the interval between pulses
is shorter than $\trelax$, then the effect of successive pulses will
accumulate, and eventually drive the transition. In Figure 5 we
show the theoretical prediction along with the respective experimental
data in the \GaTaSe\ Mott insulator. An excellent qualitative
agreement is observed, which provides a definite nontrivial
validation of our model.

In conclusion, we found universal behavior in electric field-induced insulator-to-metal transition experiments of
various prototypical narrow-gap Mott insulator compounds, \VO, \NiS\  and  \GaTaSe. These features motivated
the formulation of a resistor network model, which we
studied numerically. We obtained excellent agreement with the
experimental data, and also with a non-trivial delayed voltage
pulsing experiment predicted by the model. The numerical
simulations showed that the physical origin of the Mott resistive
transition is a dynamically directed avalanche mechanism.
Our study brings new insight to the difficult problem of the
behavior of Mott systems out of equilibrium, and is a timely
stepping-stone for research in the emerging field of Mottronic
devices.

\section*{Acknowledgments}
This work was supported by the French Agence Nationale de la Recherche through the funding of the ``NanoMott'' (ANR-09-Blan-0154-01) and ``Mott-RAM'' (ANR-2011-EMMA-016-01) projects. The authors acknowledge Prof. Aharon Kapitulnik, Prof. D. Roditchev, Dr T. Cren and Dr V. Ta Phuoc for useful discussions.

\bibliographystyle{apsrev4-1}


%

\end{document}